\documentclass[preprint]{aastex63}

\usepackage{hyperref}
\hypersetup{linkcolor=red,citecolor=blue,filecolor=magenta,urlcolor=cyan}
\received{xxx}
\revised{xxx}
\accepted{xxx}
\submitjournal{ApJ}

\shorttitle{Closed Field Lines Crossing The Coronal Holes}
\shortauthors{Huang et al.}
\begin{document}

\title{Existence of The Closed Magnetic Field Lines Crossing The Coronal 
Hole Boundaries }

%%
%% The \author command is the same as before except it now takes an optional
%% argument which is the 16 digit ORCID. The syntax is:
%% \author[xxxx-xxxx-xxxx-xxxx]{Author Name}
%%
%%
%% Note that \altaffilmark and \altaffiltext have been removed and thus 
%% can not be used to document secondary affiliations. If they are used latex
%% will issue a specific error message and quit. Please use multiple 
%% \affiliation calls for to document more than one affiliation.
%%
%%
%% Use \email to set provide email addresses. Each \email will appear on its
%% own line so you can put multiple email address in one \email call. A new
%% \correspondingauthor command is available in V6.3 to identify the
%% corresponding author of the manuscript. It is the author's responsibility
%% to make sure this name is also in the author list.
%%
%%

\correspondingauthor{Chia-Hsien Lin}
\email{chlin@jupiter.ss.ncu.edu.tw, chlin@g.ncu.edu.tw}

\author{Guan-Han Huang}
\affiliation{Department of Space Science and Engineering, \\
National Central University, \\
Taoyuan, Taiwan}

\author[0000-0002-0786-7307]{Chia-Hsien Lin}
\affiliation{Department of Space Science and Engineering, \\
National Central University \\
Taoyuan, Taiwan}

%\collaboration{1}{(AAS Journals Data Scientists collaboration)}

\author[0000-0003-4012-991X]{Luo-Chuang Lee}
\affiliation{Department of Space Science and Engineering, \\
National Central University, \\
Taoyuan, Taiwan}
\affiliation{Institute of Earth Sciences, Academia Sinica, Taipei, Taiwan}
\affiliation{State Key Laboratory of Lunar and Planetary Sciences, 
Macau University of Science and Technology, Macau, PR China}

\begin{abstract}
Coronal holes (CHs) are regions with unbalanced magnetic flux,
and have been associated with open magnetic field (OMF) structures.
However,
it has been reported that some CHs do not intersect with OMF regions.
To investigate the inconsistency,
we apply a potential-field (PF) model to
construct the magnetic fields of the coronal holes.
As a comparison, we also use a thermodynamic magnetohydrodynamic (MHD) model to
synthesize coronal images, and identify CHs from the synthetic images.
The results from both the potential-field CHs and synthetic MHD CHs
reveal that there is a significant percentage of closed field lines
extending beyond the CH boundaries and more than 50\% (17\%) of PF
(MHD) CHs do not contain OMF lines.
The boundary-crossing field lines are more likely to be found
in the lower latitudes during active times.
While they tend to be located slightly closer than 
the non-boundary-crossing ones to the CH boundaries, nearly 40\% (20\%)
of them in PF (MHD) CHs are not located in the boundary regions.
The CHs without open field lines are often
smaller and less unipolar than those with open field lines.
The MHD model indicates higher temperature variations along the
boundary-crossing field lines than the non-boundary crossing ones.
The main difference between
the results of the two models is that
the dominant field lines in the PF and MHD CHs are closed and open field
lines, respectively.
% PF and MHD results is that
%the dominant field lines are closed field lines in the PF CHs and
%are open field lines in the synthetic MHD CHs.
\end{abstract}



\section{Introduction} \label{sec:intro}
Coronal holes (CHs) are observed as the darkest regions in the coronal
images with predominantly unipolar magnetic fields at the photosphere.
They are the regions with lower coronal emissions and density
than the surrounding corona \citep[e.g.,][]{W1975SoPh,Z1977Rv,WS2004SoPh}
and an un-balanced magnetic-field distribution at
the photosphere \citep[e.g.,][]{Z1977Rv,C2009LRSP,Hofmeisteretal2017ApJ,Hofmeisteretal2019AA}.
The reconstruction of coronal magnetic fields from earlier studies
indicate that CHs are often located in the regions 
with rapidly diverging ``open'' magnetic fields
\citep[e.g.,][]{Wilcox1968SSRv, Levine1982SoPh, C2009LRSP}
which can accelerate plasma to supersonic outflows \citep{KH1976SoPh}.
``Open'' magnetic fields (OMFs) are the magnetic structures that
extend far away from the Sun.
Therefore, CHs have been considered as major source regions 
for high-speed solar wind streams \citep{Krieger1973,Krieger1974}.

However, more recent studies have shown significant 
inconsistency between 
the CHs
and the source regions of high-speed solar wind
\citep[e.g.,][]{HLL2017NatSR},
and between CHs and OMF regions
\citep[e.g.,][]{OS1999SoPh,LQLL2014ApJ,LQLL2017SoPh,Linkeretal2017ApJ,HLL2017NatSR,HLL2019ApJ}.
Part of the inconsistency can be caused by methodological inaccuracy.
For instance,
different methods of extracting dark regions
can result in 26\% variation in the unbalanced flux \citep{Linkeretal2021ApJ}.
The choice of the boundary conditions in the model employed to construct
the coronal magnetic fields can affect the topology of the determined 
OMF regions \citep{Asvestarietal2019JGRA,uncertainty,Caplanetal2021ApJ}. 
In addition to
the methodological uncertainties and inaccuracy, 
some physical effects/mechanisms are likely to also contribute to the inconsistency.
For instance,
\citet{HLL2019ApJ} show that OMF structures with higher super-radial
expansion tend to be brighter.
The analysis by \citet{HLL2019ApJ} reveals that 
some CHs (referred to as LIR in the paper)
do not overlap with OMF regions, which indicates that these predominantly
unipolar dark regions do not contain open magnetic field structures.
While it has been known that closed-field structures exist inside the CHs
\citep[e.g.,][]{WS2004SoPh, Hofmeisteretal2017ApJ,Hofmeisteretal2019AA,Heinemannetal2021SoPh2},
these structures, being bi-polar, 
cannot contribute to the observed unipolarity of the region.

The objective of current study is to investigate the source for
the absence of open magnetic structures in some CHs.
The procedure is as follows:
The first step is to construct synoptic maps of 
coronal images and magnetograms;
second step is to identify the CHs 
from the maps based on a thresholding method;
third step is to construct the coronal magnetic fields using 
Potential Field Source Surface (PFSS) model; 
fourth step is to compute the magnetic structures in the CHs
by tracing the field lines from the surface up;
and, lastly, different types of CH magnetic structures are 
identified and their characteristics examined.
To investigate how the results would be affected by the current-free
restriction in the PFSS model,
we conduct a parallel examination using
the synthetic coronal structures
distributed by Predictive Science \citep{predsci},
who computed the magnetic and thermal structures of the corona 
based on thermodynamic magnetohydrodynamics (MHD).
We first construct a set of synthetic synoptic maps of coronal emission 
using the density and temperature in the MHD model,
then identify synthetic CHs from the maps, 
and finally trace and examine the field lines in the synthetic CHs.

The observational data used for the study are described in Sec.~\ref{sec:data}.
In Sec.~\ref{sec:method},
we explain the methods for
  constructing the synoptic maps (Sec.~\ref{subsec:synoptic}),
identifying the coronal holes (Sec.~\ref{subsec:LIR}),
constructing the potential magnetic fields (Sec.~\ref{subsec:PFSS}),
MHD modeling of the corona (Sec.~\ref{subsec:MHD})
and the tracing of the field lines (Sec.~\ref{subsec:tracing}).
The results are presented in Sec.~\ref{sec:results},
discussed in Sec.~\ref{sec:discuss}, and summarized in Sec.~\ref{sec:summary}.

\section{Observational Data} \label{sec:data}
The line-of-sight (LOS)
magnetograms from Helioseismic and Magnetic Imager (HMI) and
the 193~\AA \ wavelength band images from Atmospheric Imaging Assembly (AIA),
both onboard the Solar Dynamic Observatory (SDO),
are used for the study.
The time span of the data is from June 2010 to February 2020, corresponding to
the Carrington Rotation (CR) number 2099 to 2227.

The monthly sunspot number (SSN)
data distributed by Sunspot Index and Long-term Solar Observations,
Royal Observatory of Belgium, Brussels \href{https://wwwbis.sidc.be/silso/}{(SILSO World Data Center)},
are used to evaluate the activity level of the Sun.
To determine the SSN at a specific Carrington Rotation (CR),
the monthly SSN is linearly interpolated to the time of the CR.

\section{Method} \label{sec:method}
\subsection{Synoptic Map Construction} \label{subsec:synoptic}
The LOS magnetic fields are converted into radial magnetic fields 
($B_r$)
assuming that the surface fields are radial \citep{radial1,radial2,GP2013SoPh}.
The synoptic maps of the radial fields and AIA 193 \AA \ emissions 
are constructed using the stripes of $\pm 15^\circ$ 
around the central meridian.
The size of the resulting maps is 3600 pixels in longitude ($\phi$) and 
1440 pixels in sine-latitude ($\sin \lambda$).

\subsection{Coronal Hole Identification} \label{subsec:LIR}
The coronal holes are identified as the dark regions with predominantly
unipolar magnetic field.
The identification procedure consists of three steps:
(1) extraction of dark regions from AIA 193\AA \ synoptic maps;
(2) evaluation of the stability of the dark-region boundaries, following
the stability test procedure developed by \citet{Heinemannetal2019SoPh};
and (3) evaluation of the level of unipolarity of the dark regions.

\begin{figure}
  \centering
  \includegraphics[width=.8\textwidth,clip,trim=0 4.2cm 0 7cm]{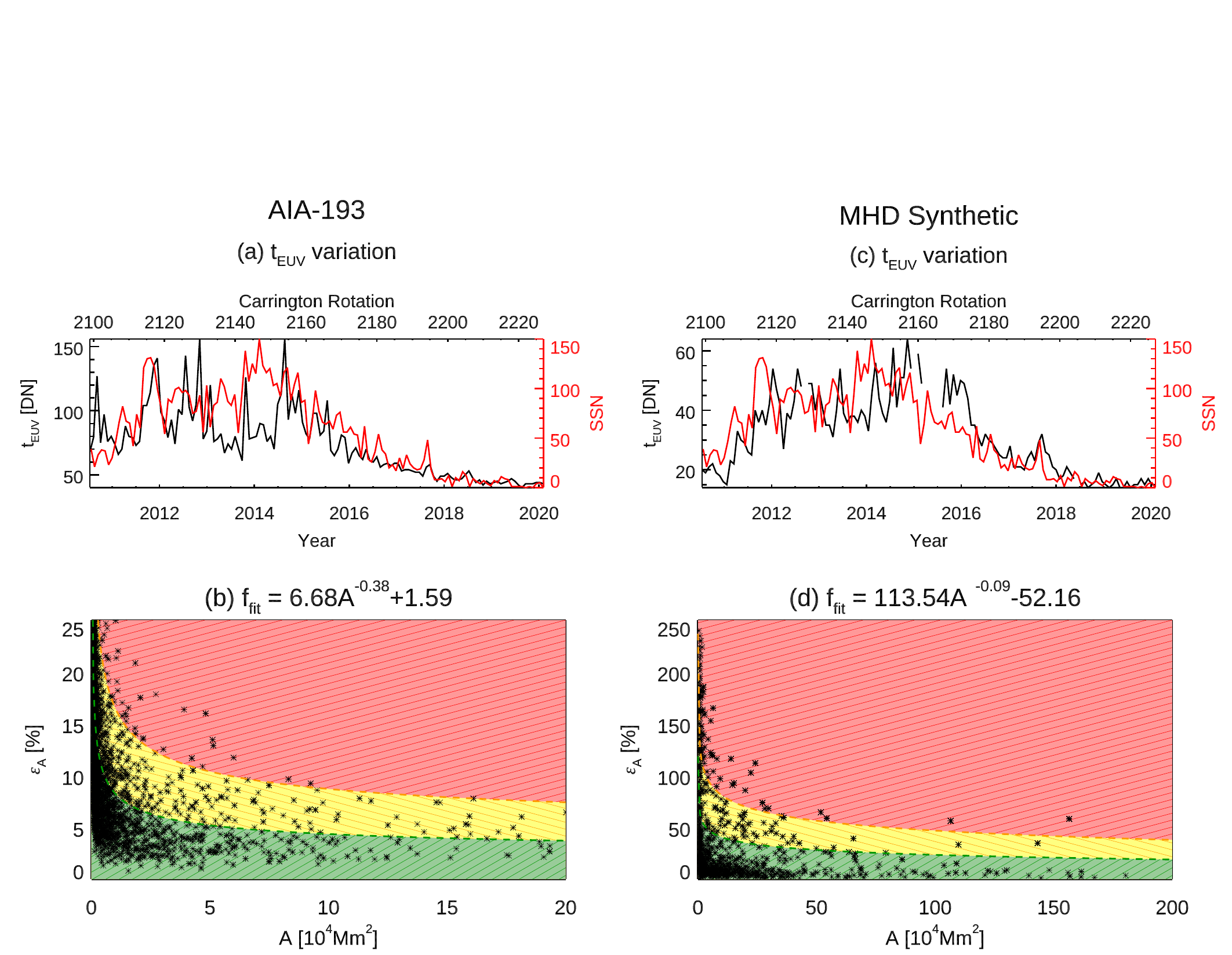}
  \caption{The upper  panels:
  the temporal variations of the optimal EUV threshold $t_\mathrm{EUV}$
  for AIA-193 maps (panel a) and MHD synthetic maps (panel c).
  The lower two panels: 
  the scatter plots of the area uncertainty $\epsilon_A$ of the dark regions 
  vs. the average dark-region area $A$ over five EUV thresholds for
  the AIA maps (panel b) and the synthetic maps (panel d).
  The best-fit curve $f_\mathrm{fit}$ is plotted as green dashed line,
  and its functional form is printed in the title of the corresponding
  panel.
  The points of high, median and low boundary stability are located in
  green, yellow, and red regions, respectively.}
  \label{fig:uncertainty}
\end{figure}

In the first step, an automated dark-region detecting procedure 
\citep{KG2009SoPh,HLL2019ApJ} is applied to AIA synoptic maps
to determine an optimal intensity threshold of 
the Extreme UltraViolet (EUV) emission, $t_{\rm EUV}$,
for each synoptic map. 
The determined $t_{\rm EUV}$ of each synoptic map
over the time period examined in this study is shown in 
Figure~\ref{fig:uncertainty}(a).
The regions that are darker than $t_{\rm EUV}$ and 
larger than 400 pixels ($\approx 470$~Mm$^2$) are extracted as dark regions.

In the second step, 
the stability of a dark region is evaluated by the variability of its area
(area uncertainty)
due to the variation of the intensity threshold \citep{Heinemannetal2019SoPh}.
The smaller the area uncertainty, the higher the stability of the dark region.
In brief, the stability evaluation is conducted as follows:
The intensity threshold is varied
from $t_{\rm EUV} - 2$ digital number (DN) to $t_{\rm EUV} + 2$ (DN),
producing five area values ($A_i$, $i=1$ to $5$) and an average area, 
$A$, for each dark region.
The area uncertainty $\epsilon_A$ of the dark region 
is subsequently computed as follows:
\begin{eqnarray}
  \epsilon_A &\equiv& \frac{{\rm max}(|A_i - A|)}{A}.
\end{eqnarray}
To determine the stability of the dark-region boundaries,
we use a power function $f(A) = \alpha A^{\beta} + \gamma $,
where $\alpha$, $\beta$ and $\gamma$ are the fitting parameters,
to fit the scatter plot of $\epsilon_A$ vs. $A$ 
(shown in Fig.~\ref{fig:uncertainty}b).
The best-fit curve $f_{\rm fit}$ is used as the bench mark to
define high, medium and low stability regions:
$\epsilon_A < f_{\rm fit}$ is high,
$f_{\rm fit} < \epsilon_A < 2 f_{\rm fit}$ is medium,
and $\epsilon_A > 2 f_{\rm fit}$ is low stability.
They correspond to the green, yellow and red regions in
Fig.~\ref{fig:uncertainty}b.

\begin{figure}
  \centering
  \includegraphics[width=0.8\textwidth,clip]{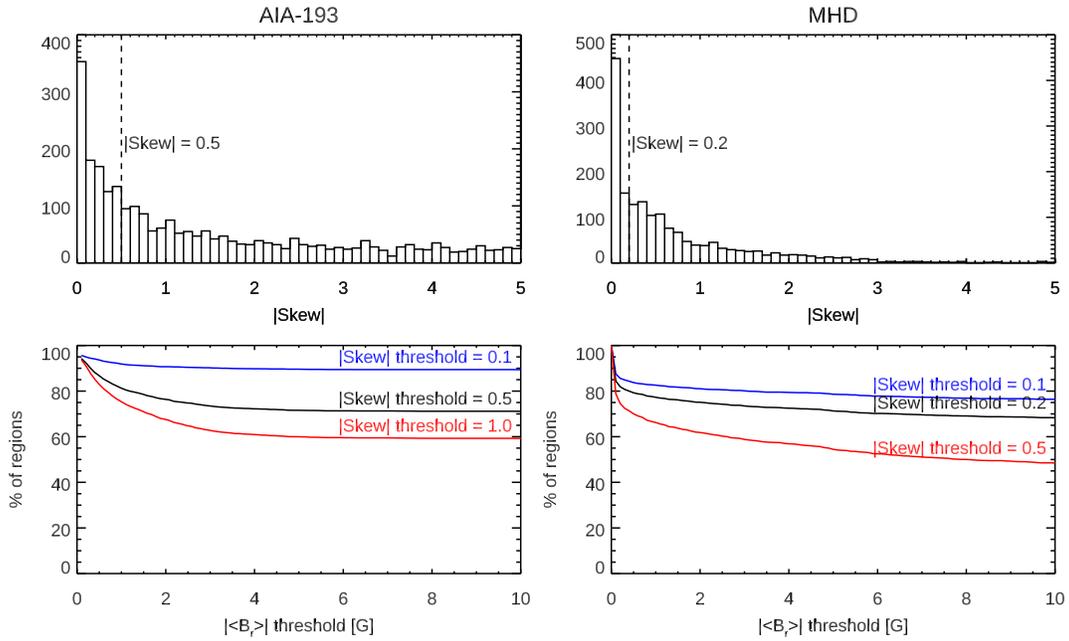}
  \caption{
           The upper left and upper right panels are 
           the histograms of skewness, $|{\rm Skew}|$,
           of the dark regions identified from
           the AIA-193 synoptic maps and 
           synthetic EUV synoptic maps, respectively .
           The lower panels show the percentage of the identified CHs
           as a function of different $|\langle B_r \rangle|$ thresholds
           for fixed $|{\rm Skew}|$ thresholds.
           Different colors represent the results of setting
           different $|{\rm Skew}|$ thresholds, as noted above 
           the corresponding curves.
           The curves for the AIA-193 and
           synthetic maps are plotted in the lower left and lower right,
           respectively.
           }
  \label{fig:ch_diff}
\end{figure}

The third step is to determine the dark regions that are predominantly 
unipolar.
A region with high level of unipolarity can be observed as a region
with the radial-field distribution highly skewed toward one polarity, or
with a large average radial field $|\langle B_r \rangle|$, or both.
Since the radial fields are measured at the photosphere,
the dark-region boundaries
are projected down to the photosphere 
by assuming that the 193 \AA \ images
represent the corona at a height of 0.02$R_\odot$ above the solar surface 
\citep{AIA_height, HLL2019ApJ}.
The histogram of the skewness of the dark regions is plotted in the upper
left panel of Fig.~\ref{fig:ch_diff}.
The x axis is $|{\rm Skew}|$, and y axis is the number of the dark regions.
The vertical line marks the location of $|{\rm Skew}| = 0.5$,
which is approximately the turning point of the histogram.
It is thus selected as the threshold for the skewness \citep{HLL2019ApJ}.
To determine the threshold for $|\langle B_r \rangle|$,
the percentage of the dark regions qualified as CHs is plotted
as a function of $|\langle B_r \rangle|$ threshold in the lower left panel
of Fig.~\ref{fig:ch_diff}.
In addition to the result of setting the skewness threshold $|{\rm Skew}|=0.5$
(black line), the
profiles of setting $|{\rm Skew}| = 0.1$ (blue) and 1.0 (red) are also plotted
for comparison.
The plot shows that the percentage of the dark regions qualified as CHs 
approaches a constant of $\approx 70\%$ after 3~Gauss.
Therefore, 3~Gauss is chosen as the threshold, and
the dark regions with $|\langle B_r \rangle| \ge 3$~Gauss
or magnetic-field skewness $|{\rm Skew}| \ge 0.5$
are considered as predominantly unipolar,
and identified as coronal holes.

Since the boundary of a dark region can be changed by the variation of 
the intensity threshold, 
$|\langle B_r \rangle|$ and $|{\rm Skew}|$ within the boundary
can also be changed. 
We consider the dark regions that are qualified as coronal holes under
all five intensity thresholds as reliable coronal holes.
The regions that only satisfy the unipolarity criteria for some
intensity thresholds are classified as uncertain coronal holes.
The coronal holes identified from the synoptic maps of
AIA 193\AA \ and HMI radial fields are referred to as CH$_{193}$ hereinafter.

\subsection{PFSS Modelling} \label{subsec:PFSS}
The global coronal magnetic field is often constructed by the Potential
Field Source Surface (PFSS) model \citep{pfss}, 
which is based on the assumptions
that the current is zero in the region of interest and that 
the tangential magnetic field vanishes at
the upper boundary, called ``source surface'',
leading to strictly radial field lines afterward.
Earlier studies \citep[e.g.,][]{pfss}
have shown that the current-free assumption
is broadly valid in most part of the corona except for the regions with
rapidly varying magnetic fields, such as sunspots and streamers.

In this study, the PFSS magnetic fields are computed using the 
Global Potential and Linear Force-Free Fields code
\citep[GLFFF;][]{JF2012SoPh},
which solves the Laplace equation by the
relaxation method.
The upper boundary, i.e., the source surface, is placed at $2.5 R_\odot$ from
the Sun center 
\citep[e.g.,][]{WS1990ApJ,OS1999SoPh}.
The upper boundary condition is that
all field lines at the source surface are assumed to be radial.
The lower boundary is at the solar surface,
and the HMI magnetic field synoptic maps are used as the lower boundary
condition.
To comply with our computing capacity,
the HMI synoptic maps (3600$\times$1400 pixels) 
are down sampled by Sifting Convolution on the Sphere \citep{RM2021ISPL}.
In brief, we first apply a low-pass Gaussian filter 
on the spherical harmonic coefficients of the maps
and then inverse transform them back
to a size of 360 pixels in longitude ($\phi$) and 
180 pixels in latitude ($\lambda$).
The detail of the down sampling procedure is described in Appendix.

\subsection{MHD Modelling} \label{subsec:MHD}
The data of MHD modelled solar corona for the time period studied in this work
are from \href{http://www.predsci.com/mhdweb/data_access.php}{Predictive Science},
who computed the magnetic and thermal structures of the corona
using their
Magnetohydrodynamic Algorithm outside a Sphere (MAS) code,
which solves the full set of thermodynamic MHD equations.
The plasmas are assumed to be collisional below
$10 R_\odot$  and collisionless beyond $10 R_\odot$.
The energy equation in the code includes the contributions from
coronal heating, dissipation and radiative loss.

Using the electron density $n_e$ and temperature $T_e$ in the MAS data,
we compute the synthetic EUV emission maps at 193 \AA \ 
wavelength by the following equation:
\begin{equation}
   I_{193} = \int n_e^2(z) f_{193}(T_e(z))\mathrm{d}z,
\label{eqn:EUV}
\end{equation}
where $n_e(z), T_e(z)$ are the electron number density and the
electron temperature along the line of sight $z$, and
$f_{193}(T_e)$ is the AIA-193 temperature response function.
The coronal holes for the MHD model are identified
following the same procedure described in Section~\ref{subsec:LIR}:
(1) the dark regions are extracted from the synthetic EUV synoptic maps
by intensity thresholding,
(2) their stability is evaluated by the stability test,
and 
(3) a set of unipolarity criteria is applied to identify the coronal holes.
The unipolarity criteria are
determined by the same procedure described in Sec.~\ref{subsec:LIR}:
The histogram of $|{\rm Skew}|$
is plotted in the upper right panel of Fig.~\ref{fig:ch_diff}.
The vertical line in the upper right panel marks the location $|{\rm Skew}| = 0.2$,
which is where the slope of the histogram shows the sharpest change.
The percentage of the qualified CHs as a function of 
$|\langle B_r \rangle|$ is shown in the lower right panel of 
Fig.~\ref{fig:ch_diff}.
The result of setting the skewness threshold to 0.2 is plotted in black,
and compared with the results of setting the threshod to 0.1 (blue)
and 0.5 (red).
The plot shows that the percentage of the qualified CHs approaches
a constant of $70\%$ after $|\langle B_r \rangle| = 2$~G.
Therefore, $|\langle B_r \rangle| = 2$~G is selected as the threshold for
$|\langle B_r \rangle|$.
In short, based on these plots,
we determine the unipolarity criteria for the
synthetic maps to be
$|{\rm Skew}| \ge 0.2$ or 
$|\langle B_r \rangle| \geq 2.0$~Gauss.
The CHs identified from the synthetic synoptic maps will be referred to
as synthetic MHD CHs or CH$_{\rm MHD}$,
to be distinguished from CH$_{193}$,
the coronal holes identified from the AIA and HMI synoptic maps.

\subsection{Field Line Tracing} \label{subsec:tracing}
The magnetic field lines are computed by
tracing the field lines from the surface upward.
The ``open'' magnetic field lines in a PFSS magnetic field are defined
as the field lines reaching 
the source surface, which is $2.5 R_\odot$ from the Sun center.
To define the OMF lines in the MHD magnetic fields, 
we computed
the distance at which the thermal pressure of the MHD model
becomes larger than the tangential
magnetic pressure.
The result indicates that except for CR2157, 
``open'' MHD field lines can also be defined as the field lines
reaching $2.5 R_\odot$ from the Sun center.

After the field line tracing, the field lines that are
rooted in the reliable coronal holes are considered as reliable field lines
to be used for the analysis of CH magnetic structures.
Other field lines are categorized as uncertain field lines and discarded.

\section{Results} \label{sec:results}

\subsection{Reliable coronal holes and field lines}
The temporal variations of the optimal intensity thresholds $t_{\rm EUV}$
in AIA-193 synoptic maps and in MHD synthetic EUV maps are
plotted as black lines in Fig.~\ref{fig:uncertainty}a and b, respectively.
The overplotted red lines are the SSN, and follow the y-axis on the right side.
The panels show that the $t_{\rm EUV}$ variations of both AIA and synthetic 
maps are large before 2016,
when the solar activity is higher,
and become small with $t_{\rm EUV}$ approaching a constant after 2016, when
the Sun becomes quieter.

%% scatter plots -- stability
The scatter plots of $\epsilon_A$ vs. $A$ are shown in
Fig.~\ref{fig:uncertainty}b and d for the dark regions identified from
the AIA-193 synoptic maps and synthetic EUV maps, respectively.
The best-fit curves to the scatter plots are plotted
as green dashed curves, and the functional forms 
printed in the titles of corresponding panels.
The points corresponding to high, medium and low stabilities are located in
green, yellow and red sections, respectively. 
Among the 3210 dark regions extracted from the AIA-193 synoptic maps,
1894 ($\approx 59.0\%$) regions are categorized as high stability,
1133 ($\approx 35.2\%$) as medium stability,
and 183 ($\approx 5.7\%$) as low stability.
From the synthetic EUV synoptic maps, 1506 dark regions are identified,
of which 924 ($\approx 61.4\%$) are categorized as high stability,
308 ($\approx 20.4\%$) as medium stability,
and 274 ($\approx 18.2\%$) as low stability.

%% scatter plots -- area uncertainty magnitude
The y-axis of the scatter plots
show that the area uncertainties for AIA-193 dark regions 
are all less than 25\% of their average areas
while $\epsilon_A$ for the dark regions
extracted from the synthetic EUV maps
can be higher than 200\%, indicating that
the dark regions extracted from the AIA maps are more
stable than those from the MHD synthetic maps.
The large $\epsilon_A$ for the synthetic dark regions is because
the synthetic EUV maps are smoother, i.e.,
the intensity gradient is smaller.
As a result, a same change in $t_{\rm EUV}$ from -2 to 2 DN results in
larger area change. 

\begin{figure}
  \centering
  \includegraphics[width=0.8\textwidth,clip]{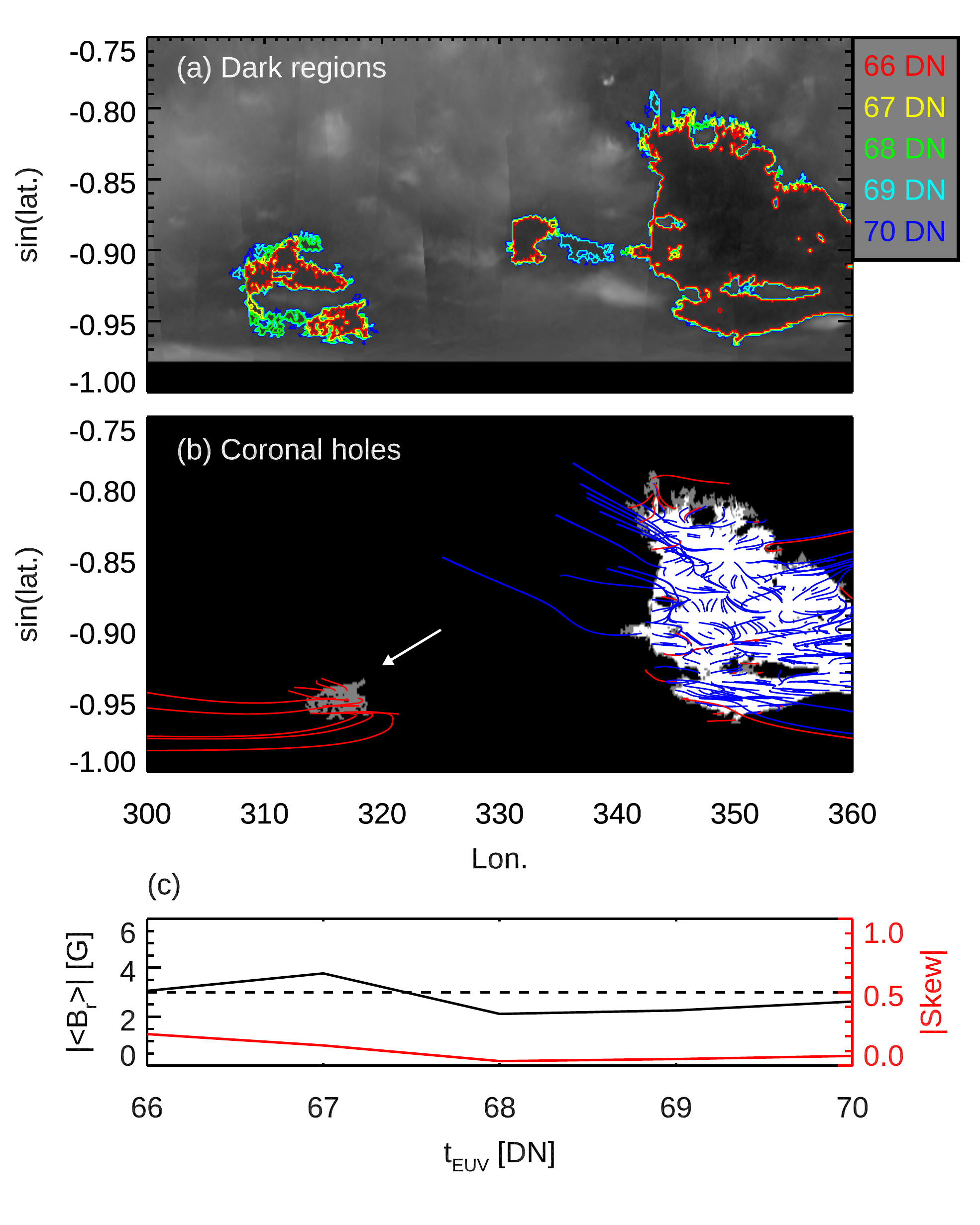}
  \caption{An example 
           to show the effect of the intensity thresholds on the 
           dark region boundaries and the unipolarity levels.
          (a) The background is the AIA-193 synoptic map of CR2099.
%              The optimal EUV threshold for CR2099 is $t_{\rm EUV} = 68$~DN.
              Contours of different colors mark the boundaries of
              the dark regions determined using different intensity thresholds:
              66, 67, 68, 69, and 70~DN, as indicated.
%           The lower-left region is an example of low stability,
%           and the larger region at the right side is a high-stability region.
%
         (b) The panel shows the dark regions 
             which satisfy the unipolarity criteria
             ($|\langle B_r \rangle| \ge 3$~G or $|{\rm Skew}| \ge 0.5$)
             and are identified as coronal holes. 
             The white and gray regions represent the reliable and uncertain 
             coronal hole areas, and
             the blue and red lines trace 
             the reliable and uncertain field lines, respectively.
         (c)  This panel shows the variations of $|\langle B_r \rangle|$ 
             (black line) and $|{\rm Skew}|$ (red line) 
             of the gray region indicated by the white arrow in panel~(b)
             when the intensity threshold is varied from 66 to 70~DN.
             The unipolarity criteria is plotted as a horizontal dashed line.
             The region is disqualified as a coronal hole
             when the intensity threshold is higher than $\approx 67.5$~DN.
      }
\label{fig:ch_reli}
\end{figure}

\begin{figure}
  \centering
  \includegraphics[width=0.8\textwidth,clip]{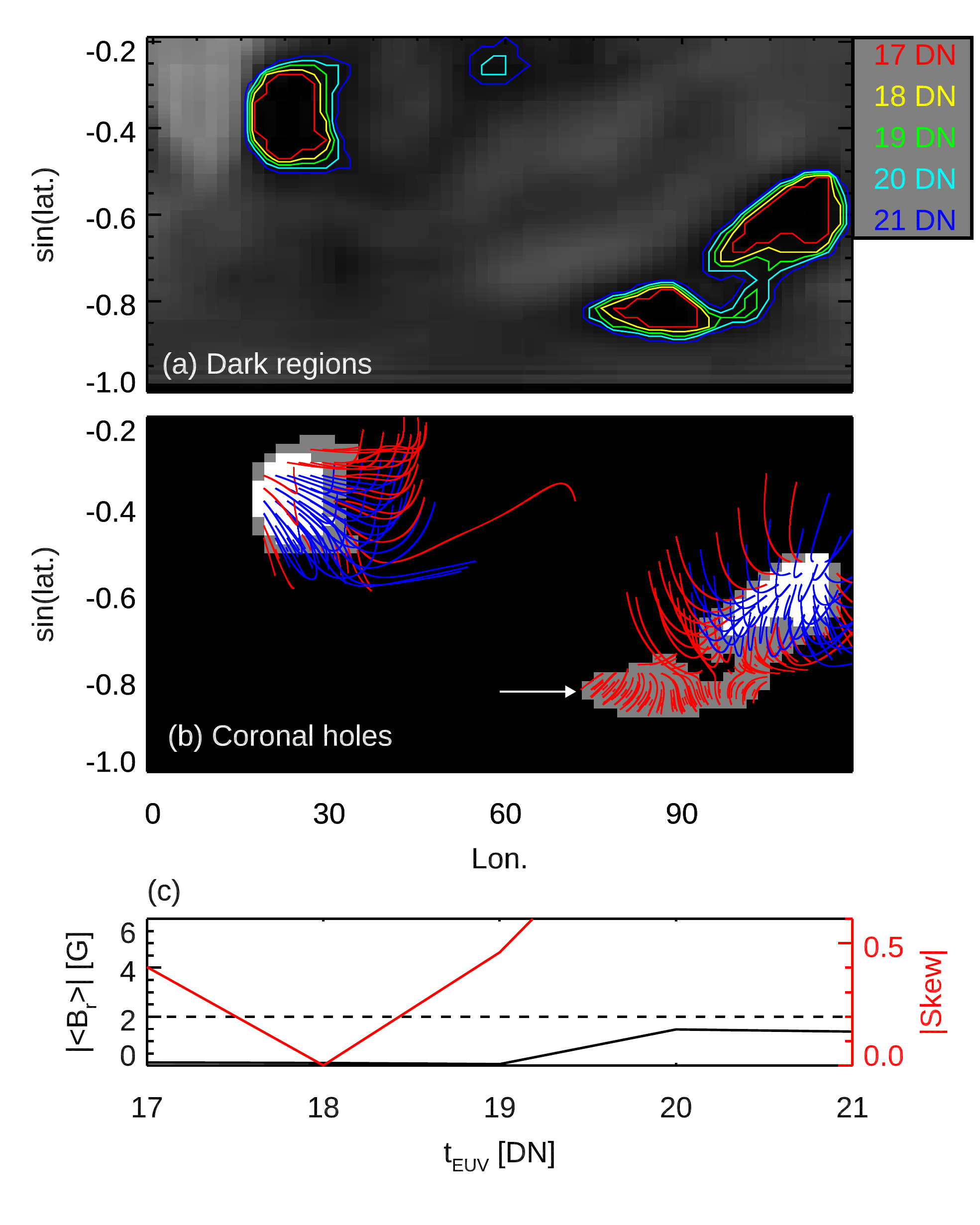}
  \caption{An example
           to show the effect of intensity thresholds on the
           dark region boundaries and the unipolarity levels.
          (a) The background is the synthetic EUV synoptic map
              of CR2100.
              Contours of different colors mark the boundaries of
              the dark regions determined using different intensity thresholds:
              17, 18, 19, 20, and 21~DN, as indicated.
         (b) The panel shows 
             the dark regions which satisfy the unipolarity threshold
             ($|\langle B_r \rangle| \ge 2$~G or $|{\rm Skew}| \ge 0.2$)
             and are identified as coronal holes.
             The white and gray regions represent the reliable and uncertain
             coronal hole areas, and
             the blue and red lines trace
             the reliable and uncertain field lines, respectively.
         (c) 
             This panel shows the variations of $|\langle B_r \rangle|$
             (black line) and $|{\rm Skew}|$ (red line)
             of the gray region indicated by the white arrow in panel~(b)
             when the intensity threshold is varied from 17 to 21~DN.
             The unipolarity criteria
             is plotted as a horizontal dashed line. 
             The region does not pass the unipolarity criteria
             when the intensity threshold $t_{\rm EUV}$ 
             is approximately $18$~DN 
             (or in the range of 17.5~DN $< t_{\rm EUV} <$~18.4~DN).
      }
\label{fig:lir_reli}
\end{figure}

%\subsection{Reliability of coronal holes}
%% reliability of coronal holes
Large area uncertainty 
can lower the reliability of the identified coronal holes.
This is because the distribution of the magnetic field within the boundary is 
more likely to change from 
predominantly unipolar to non-unipolar, or vice versa,
if the enclosed area is changed by a large percentage.
Fig.~\ref{fig:ch_reli} and Fig.~\ref{fig:lir_reli} show such examples
found in the AIA-193 synoptic map of CR2099 and synthetic EUV map of CR2100,
respectively.
The background images in Fig.~\ref{fig:ch_reli}a and Fig.~\ref{fig:lir_reli}a
are the AIA-193 map and synthetic EUV map,
respectively.
The contours of different colors mark the 
boundaries of the dark regions determined by different intensity thresholds,
as indicated next to the two panel (a).
Both panels contain 
both high-stability dark region(s), 
of which the variation of 
intensity threshold leads to a relatively small area variation,
and low-stability regions, of which the same threshold variation
results in significant boundary changes.

By comparing 
the background images of
Fig.~\ref{fig:lir_reli}a and Fig.~\ref{fig:ch_reli}a,
we can see that the synthetic EUV map (Fig.~\ref{fig:lir_reli}a) is
smoother than the AIA-193 map (Fig.~\ref{fig:ch_reli}a),
and therefore
the area uncertainty of the former is larger than that of the latter,
as pointed out in the discussion of $\epsilon_A$ vs. $A$ scatter plots.

Fig.~\ref{fig:ch_reli}b and Fig.~\ref{fig:lir_reli}b
show the coronal holes identified from the dark regions 
in Fig.~\ref{fig:ch_reli}a and Fig.~\ref{fig:lir_reli}a, respectively. 
The dark regions that do not qualify as coronal holes under any intensity threshold
are not shown.
The white and gray areas 
correspond to
the reliable and uncertain coronal holes,
and blue and red lines the reliable and uncertain field lines,
respectively.
The gray region indicated by the white arrow in each figure is
an uncertain coronal hole whose 
average magnetic field and skewness are plotted as a function of
intensity threshold in Panel (c).
The dashed horizontal line in each Panel (c) marks the threshold of
unipolarity 
($|\langle B_r \rangle| \ge 3$~G or $|{\rm Skew}| \ge 0.5$ 
for Fig.~\ref{fig:ch_reli}c, and 
$|\langle B_r \rangle| \ge 2$~G or 
$|{\rm Skew}| \ge 0.2$ 
for Fig.~\ref{fig:lir_reli}c).
The $|\langle B_r \rangle|$ and $|{\rm Skew}|$
of the indicated gray regions are plotted as black and red lines, 
respectively.
The x axis is the intensity threshold $t_{\rm EUV}$.
$|\langle B_r \rangle|$ follows the y axis on the left, and
$|{\rm Skew}|$ the y axis on the right.

Fig.~\ref{fig:ch_reli}c indicates that 
both $|{\rm Skew}|$ and $|\langle B_r \rangle|$ become 
lower than the unipolarity thresholds
when $t_{\rm EUV}$ is greater than approximately $67.5$~DN, 
disqualifying the region as a CH.
Fig.~\ref{fig:lir_reli}c indicates that the indicated region in 
Fig.~\ref{fig:lir_reli}b is disqualified as a CH when $t_{\rm EUV}$ is 
approximately 18 (17.5~DN $< t_{\rm EUV} < 18.4$~DN),
when both $|{\rm Skew}|$ and $|\langle B_r \rangle|$ are lower
than the unipolarity thresholds.
Since they are not qualified as coronal holes for all five intensity
thresholds, 
these two regions are categorized as uncertain coronal holes.
In contrast, 
the white regions in panel (b) of 
Fig.~\ref{fig:ch_reli} and Fig.~\ref{fig:lir_reli}
satisfy the unipolarity criteria for all five intensity thresholds.

The field lines that are rooted in the reliable CHs are considered as
reliable field lines. In total, 
81.8\% 
of the field lines in CH$_{193}$ 
and 
61.8\% 
of the field lines in CH$_{\rm MHD}$
are identified as reliable field lines.
All CHs and field lines discussed in the rest of the paper are
the reliable CHs and reliable field lines, 
and ``reliable'' will be dropped for brevity.

\subsection{Boundary-crossing and non-boundary-crossing field lines}
\label{sec:res_cross}
The examination of the field lines reveals that
in addition to the well-known open magnetic field lines and
closed field lines with both foot points located in the same CH,
there are closed field lines that cross the boundary of the CH
and end in either a bright region or a different CH.
For conciseness,
we use ``1'' and ``0'' to indicate the field-line foot points
located inside and outside a CH, respectively, 
and abbreviate OPen and CLose as OP and CL.
Therefore, OPen field lines will be referred to as OP1,
and CLosed field lines with one foot point ending in a non-CH region
will be referred to as CL10.
To distinguish the CLosed field lines with both foot points rooted in
a same CH and those connecting two different CHs,
the former is referred to as normal CLosed field lines CL11n,
and the latter as abnormal CLosed field lines CL11a.

\begin{figure}
  \centering
  \includegraphics[width=.9\textwidth,clip]{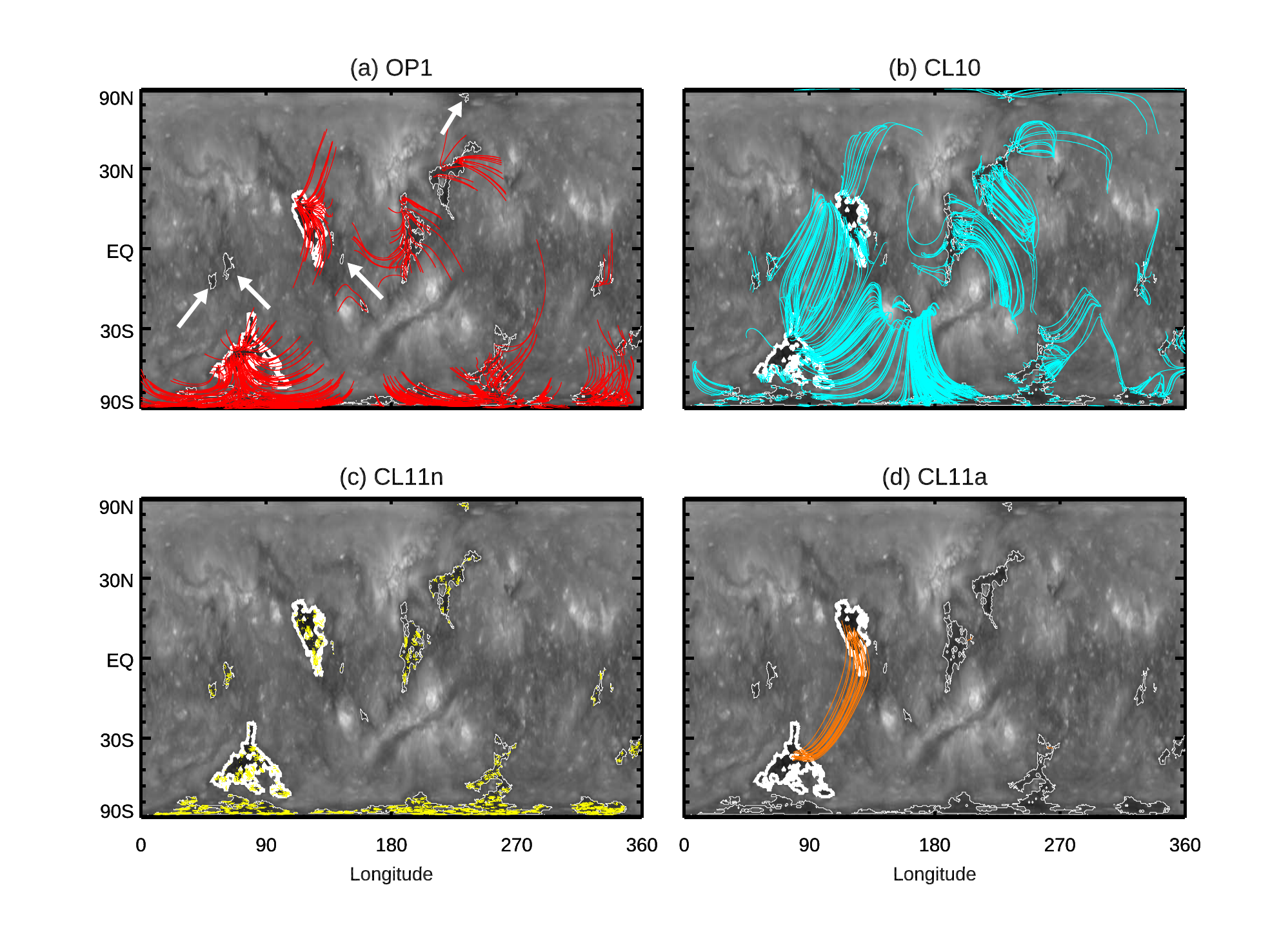}
  \caption{CR2105 reliable CH magnetic field lines computed from the PFSS model.
           The background is AIA 193 \AA \ synoptic map.
           The white contours mark the boundaries of CH$_{193}$. 
           The arrows in panel (a) point to the CH$_{193}$ with no 
           open field lines.
           Different field line types
           are plotted in different panels in different colors, 
           as indicated. 
           The percentages of their numbers in this Carrington Rotation are:
           (a) OP1 (open field lines): 20.3\%;
           (b) CL10 (closed field lines extending to  bright regions):
               27.5\%;
           (c) CL11n (closed field lines confined in the same CH$_{193}$):
               51.8\%;
           (d) CL11a (closed field lines connecting two CH$_{193}$):
               0.5\%.
          }
  \label{fig:pfss_2105}
\end{figure}

The results of CR2105 are shown in Fig.~\ref{fig:pfss_2105} as an example.
The field lines are computed from the PFSS magnetic fields.
The background in all panels is the AIA 193 \AA \ synoptic map projected
to the solar surface, 
assuming that the height of the AIA 193 images is at $0.02 R_\odot$.
The white contours mark the identified CH$_{193}$,
and different types of field lines are plotted in different colors
in separate panels, 
 as indicated in the panel titles.
Although OP1 (panel a) and CL10 (panel b) appear to be the dominant
types for this Carrington Rotation,
they occupy less than 21\% and 28\% of the total number of field lines
in this Carrington Rotation, respectively.
The most numerous type is actually CL11n (panel c). 
CL11n, while being shorter than the other three field line types,
can be found in all CHs, and accounts for nearly
52\% %(2563 out of 4951) 
of the field lines in CR2105.

Comparing panel (a) and other panels, it can be seen that
four CHs (indicated by white arrows) in this CR contain no OP1 and only
closed field lines.
CHs with such magnetic structures do not intersect with OMF regions,
and would not be the source regions for high-speed solar wind streams.
A close examination of panel (b) reveals that 
some brighter ends of CL10 are found in the neighborhoods of active regions
and just outside the edges of another CH.
CL11a, shown in panel (d), is usually the rarest among the four types.
In this Carrington Rotation, they account for only 0.5\% of the total number of 
the field lines.
In this specific CR,
they are seen to connect two CHs (marked in thicker contours) 
that are located in different hemispheres.
Comparison with panel (b) shows that these two CHs are also connected by
several CL10.

\begin{figure}
  \centering
  \includegraphics[width=.9\textwidth,clip]{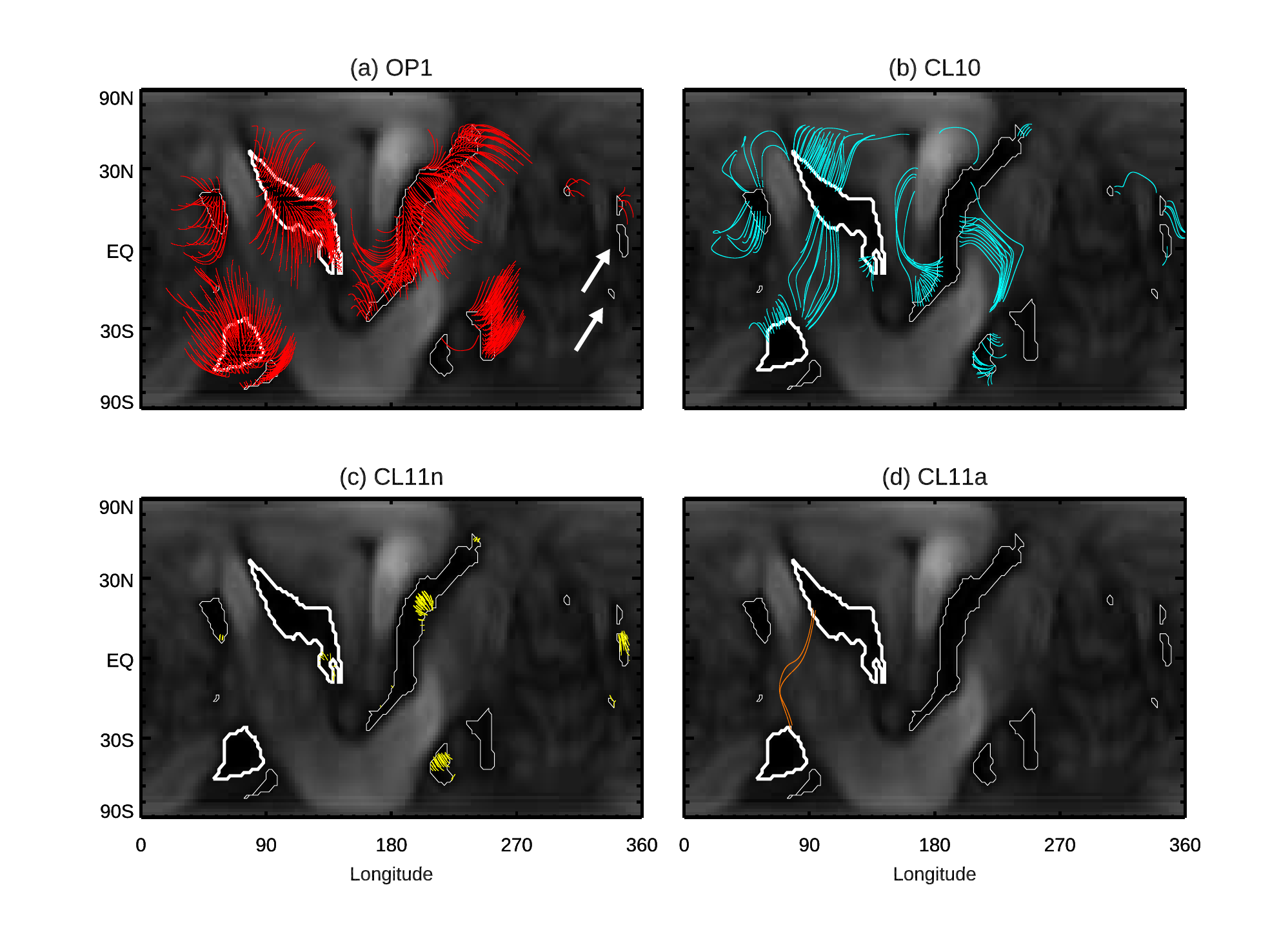}
  \caption{CR2105 reliable CH magnetic field lines computed from the MHD model. 
           The background is the synthetic EUV synoptic map.
           The white contours mark the boundaries of CH$_{\rm MHD}$. 
           The arrows in panel (a) point to the CH$_{\rm MHD}$ with no 
           open field lines.
           Different field line types are plotted in different panels in 
           different colors, as indicated. 
           The percentages of their numbers in this Carrington Rotation are
           (a) OP1 (open field lines): 73.1\%; 
           (b) CL10 (closed field lines extending to bright regions): 15.6\%;
           (c) CL11n (closed field lines confined in the same CH$_{\rm MHD}$): 
               11.1\%;
           (d) CL11a (closed field lines connecting two CH$_{\rm MHD}$): 0.2\%.
          }
  \label{fig:mhd_2105}
\end{figure}

The field lines computed from the MHD model of the same Carrington Rotation,
CR2105, are plotted in Figure~\ref{fig:mhd_2105} in the same format as
Figure~\ref{fig:pfss_2105}.
The background in all panels is the synthetic EUV map
and the white contours are the identified CH$_{\rm MHD}$.
The synthetic map and CH$_{\rm MHD}$ are qualitatively similar to,
albeit spatially less resolved than, 
the observed AIA map and CH$_{193}$ in Figure~\ref{fig:pfss_2105}.

Figure~\ref{fig:mhd_2105} shows that the coronal holes from the MHD model 
contain all four types of the field lines:
OP1 (panel a) is the most numerous field line type, accounting for
$\approx 73\%$ of the total field lines in this CR,
and CL11a (panel d) is the rarest type ($\approx 0.2\%$ of the total number).
Despite being the most numerous, OP1 does not exist in all CH$_{\rm MHD}$.
In this Carrington Rotation, there are two CH$_{\rm MHD}$ without
any OP1, as indicated by the arrows.
CL11n, which was the most numerous type in Figure~\ref{fig:pfss_2105},
constitutes only $\approx 11\%$  
of the total field lines in the CH$_{\rm MHD}$
of CR2105, and does not exist in all CH$_{\rm MHD}$.
Similar to the CL10 in the PFSS magnetic fields,
the brighter ends of CL10 (panel b) in the MHD model are also found in the 
neighborhoods of active regions and near the boundary of another coronal hole.

\subsection{Coronal holes without open magnetic field lines}
\label{sec:res_noOP1}
% Area
As revealed in Sec.~\ref{sec:res_cross},
there are coronal holes without open magnetic field lines.
They constitute more than 50\% of all CH$_{193}$,
and more than 17\% of all CH$_{\rm MHD}$.

The CHs without OP1 are on average smaller than 
those with OP1.
For CH$_{193}$,
the average area of the former is $\approx 0.46 \times 10^4$~Mm$^2$
and is $\approx 3.42 \times 10^4$~Mm$^2$ for the latter.
For CH$_{\rm MHD}$, 
the values are $\approx 0.85 \times 10^4$~Mm$^2$ (without OP1) and
$\approx 12.79 \times 10^4$~Mm$^2$ (with OP1).

% Magnetic-field distribution
To examine their magnetic fields,
the histograms of $|\langle B_r \rangle|$ and $|{\rm Skew}|$ 
of the coronal holes with and without OP1 are plotted in Fig.~\ref{fig:op1Br},
and the averages are printed in the corresponding panels.
The upper four panels are the histograms of $|\langle B_r \rangle|$
and the lower four the histograms of $|{\rm Skew}|$.
The histograms for CH$_{193}$ and CH$_{\rm MHD}$ are placed in the
left and right panels, respectively.

Compared with the $|\langle B_r \rangle|$ histograms of
the coronal holes with open magnetic field lines (panel a and e),
the histograms of the CHs without OP1 (panel b and f) 
are more concentrated near lower $|\langle B_r \rangle|$
with fewer cases at the high $|\langle B_r \rangle|$ tail.
This is especially prominent in panel f, with more than 50\% of CH$_{\rm MHD}$
concentrated at small $|\langle B_r \rangle|$ ($< 1$~G).
The high average $|\langle B_r \rangle|$ in panel f is caused by a few
outliers with very large $|\langle B_r \rangle|$ ($> 30$~G).
For the histograms of $|{\rm Skew}|$, the peaks of the distributions of
the CHs with and without OP1 are at similar locations.
However, the CHs without OP1 (panel d and h) have fewer cases than
the CHs with OP1 (panel c and g) at the tails of high $|{\rm Skew}|$.

In short,
these histograms indicate that, for both CH$_{193}$ and CH$_{\rm MHD}$,
the coronal holes without open magnetic field lines are less likely to have
large $|\langle B_r \rangle|$ and large skewness.
In other words, CHs without OP1 are less unipolar than the CHs with OP1.

\subsection{Comparison of different types of field lines}
\subsubsection{Distance to the boundary}
The distance of a field-line foot point to the CH boundary
is computed as the shortest arc distance (distance on a sphere) 
between the foot point and all boundary points of the CH.
Small bright spots within the CH are filled before the computation.
To compare the distance-to-the-boundary, $d$, of different field lines
in different CHs,
we define normalized distance as $d/d_{\rm max}$,
where $d_{\rm max}$ is the maximum of all $d$ in a same CH.
The cumulative probabilities of the normalized distances, $d/d_{\rm max}$,
of different field-line types are plotted in Fig.~\ref{fig:db_sph}. 
The x axis is $d/d_{\rm max}$, and y axis is the cumulative probability.
The results of the PFSS magnetic fields and MHD magnetic fields
are placed in the left and right columns, and
different field-line types are shown in different rows,
as noted in the panel titles.

The plots show that 
the initial slopes of the boundary-crossing field lines extending to a
bright region (CL10) and to another coronal hole (CL11a)
are slightly steeper than the slopes of the field lines confined in a same CH,
indicating that the boundary-crossing field lines
are located slightly closer to the boundaries than those confined in a CH.
For instance,
if we choose $d/d_{\rm max} \le 0.3$ as the boundary region,
the cumulative probabilities indicate that
slightly more than 60\% (80\%) of CL10 and CL11a and slightly less than
60\% (60\%) of OP1 and CL11n in CH$_{193}$ (CH$_{\rm MHD}$) 
are located in this boundary region.
On the other hand, this also indicates that 
nearly 40\% (20\%) of the boundary-crossing field lines in CH$_{193}$ 
(CH$_{\rm MHD}$) are not located in the boundary regions.

The dashed lines in the panels point to the normalized distances at which
the cumulative probability reaches 0.9, which means 90\% of field lines
are within this normalized distance from the nearest boundary.
For the PFSS magnetic field, 90\% of OP1, CL10, CL11n and CL11a are
located within 0.65, 0.59, 0.62 and 0.58 normalized distance
from the CH boundary, respectively. 
For the MHD magnetic field, the distances are 0.68, 0.50, 0.69 and 0.43.
The plot and these numbers indicate that, statistically, 
the boundary-crossing field lines,
i.e., CL10 and CL11a, are located closer to the boundaries
than the non-boundary-crossing ones are.
The difference is larger in the MHD magnetic fields than in the PFSS
magnetic fields.

\subsubsection{Temperature distribution along the field lines}
\begin{figure}
  \centering
  \includegraphics[width=.9\textwidth,clip,trim=0 10cm 0 0cm]{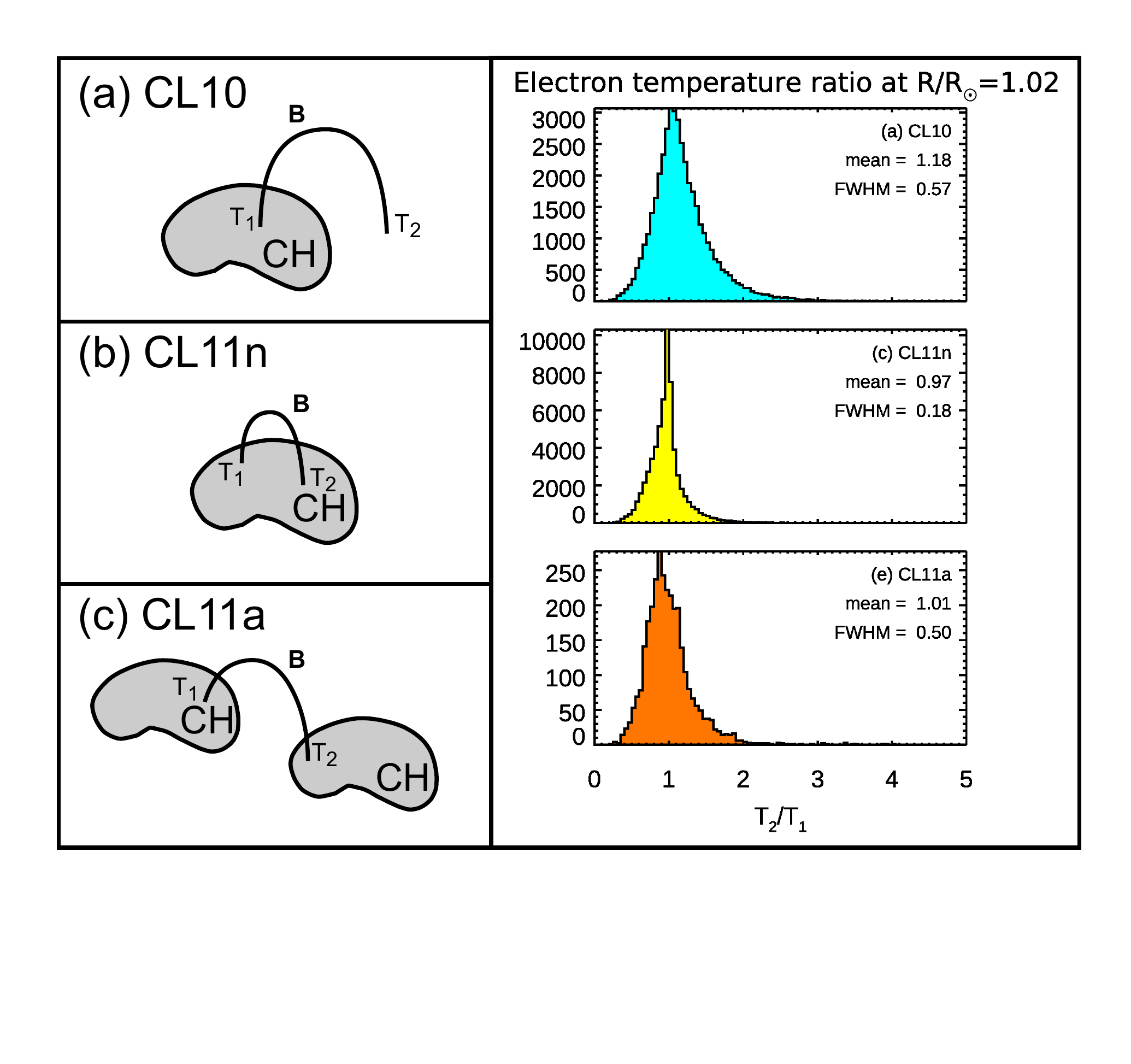}
  \caption{
           The histograms of the temperature ratio, $T_2/T_1$, 
           of the two foot points of CL10, CL11n and CL11a in CH$_{\rm MHD}$.
           For CL10, $T_2$ is the temperature of brighter foot point.
           For CL11n and CL11a, $T_2$ is the temperature of the lower-latitude
           foot point.
          }
  \label{fig:Tratio}
\end{figure}

To investigate whether any heat imbalance may exist in the
closed field lines connecting different regions,
we use the electron temperature $T_e$ in the MAS data to calculate 
the ratio of the temperatures, $T_2/T_1$, of the two foot points at
1.02 $R_\odot$.
For CL10, 
$T_2$ is defined as $T_e$ of the bright foot point
and $T_1$ that of the dark foot point.
For CL11a and CL11n, 
$T_2$ is the temperature of the lower-latitude foot point
and $T_1$ that of the higher-latitude foot point.
The histograms of $T_2/T_1$ are shown in Figure~\ref{fig:Tratio}
along with the sketches of these field line types.
The mean and Full Width at Half Maximum (FWHM) of
each distribution are printed in respective panel.
The distribution of CL10 shows that the mean $T_2/T_1$ is larger than 1.
This indicates that, on average, the brighter foot point is hotter than
the dark foot point.
FWHM of the three panels shows that
the distribution of CL11n is much narrower than that of CL10 and CL11a,
indicating that 
the field lines connecting to different regions (i.e., CL10 and CL11a) 
are more likely 
to be non-isothermal than those confined in a same region (i.e., CL11n).
This is consistent with the finding by \citet{Heinemannetal2021SoPh1},
who reported that the temperature inside the coronal hole
boundaries tends to be uniform.
Although $\langle T_2/T_1 \rangle_{\rm CL11a} = 1.01 $
and $\langle T_2/T_1 \rangle_{\rm CL11n} = 0.97 $
may seem to indicate that the field lines connecting two different CHs
are, on average, slightly more isothermal than those rooted in the same CHs,
the difference is too small to be statistically significant.

\subsubsection{Dependence on latitude and solar activity}
\begin{figure}
  \centering
  \includegraphics[width=.9\textwidth]{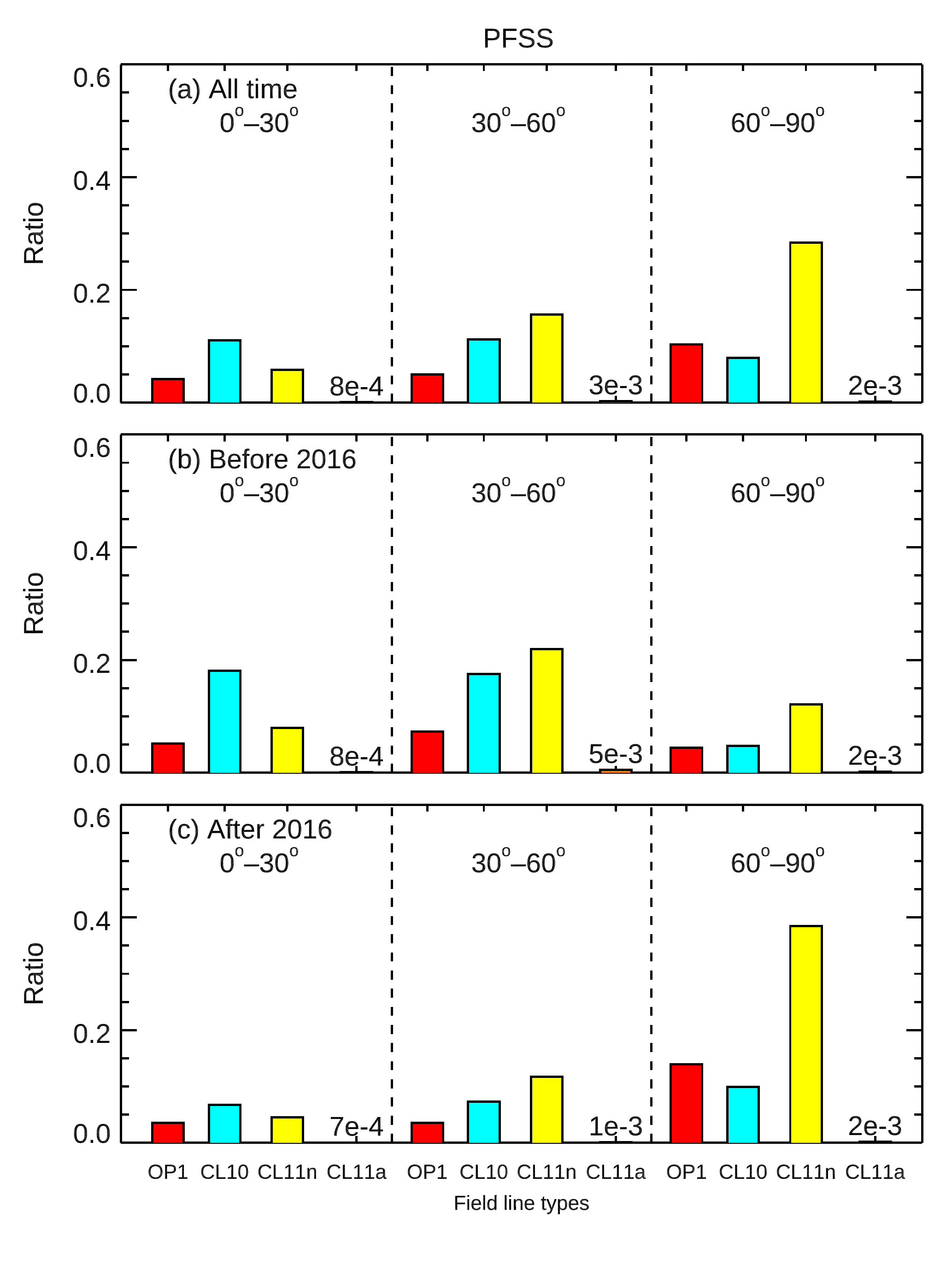}
  \caption{
           The ratio distribution 
           of OP1 (red bars), CL10 (cyan bars),
           CL11n (yellow bars), and CL11a (orange bars) in different
           latitudes and during different solar activity levels
           for the PFSS magnetic fields (CH$_{193}$).
           The panels from top down: All time, Before 2016 (Active time)
           and After 2016 (Quiet time).
           In each panel, left to right: low-, mid- and high-latitude regions.
           The height of bar corresponds to the ratio in that panel
           (i.e., the total of the twelve bars in each panel is 1).
          }
  \label{fig:fptype_lat_aq_glfff}
\end{figure}

\begin{figure}
  \centering
  \includegraphics[width=.9\textwidth]{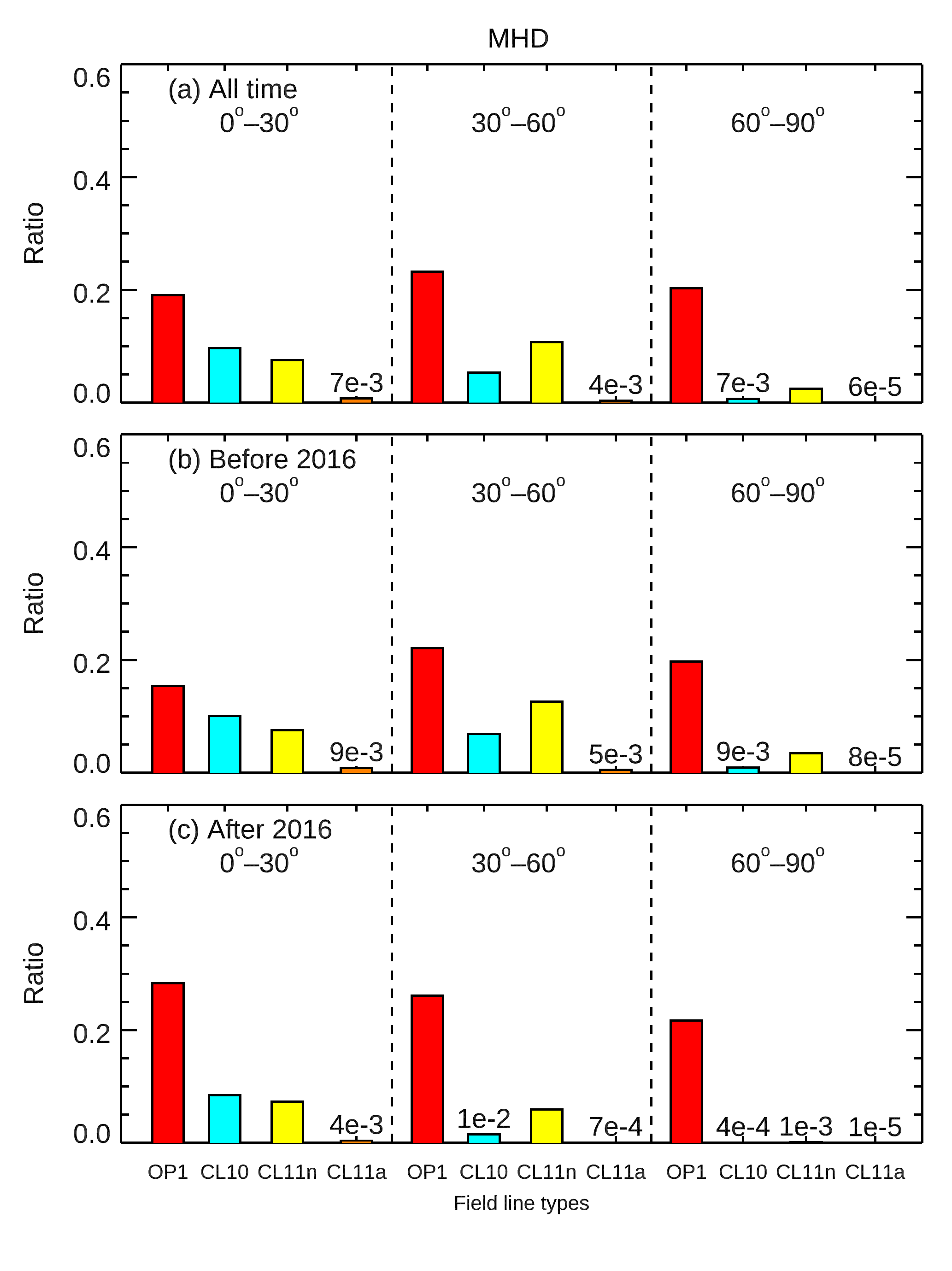}
  \caption{
           The ratio distribution 
           of different field-line types for the MHD
           magnetic fields (CH$_{\rm MHD}$) plotted in
           the same format as in Fig.~\ref{fig:fptype_lat_aq_glfff}.
           OP1, CL10, CL11n, and CL11a are represented by red, cyan, yellow,
           and orange bars, respectively.
           The panels from top down: All time, Before 2016 (Active time)
           and After 2016 (Quiet time).
           In each panel, left to right: low-, mid- and high-latitude regions.
          }
  \label{fig:fptype_lat_aq_lirmhd}
\end{figure}

To examine whether
the occurrence of different field-line types may be related to the
latitudes and solar activity,
we plot the ratio 
distributions of OP1, CL10, CL11a, and CL11n
at different latitudes and different activity levels.
The results 
from the PFSS (i.e., CH$_{193}$) and MHD magnetic fields (i.e., CH$_{\rm MHD}$)
are shown in Figure~\ref{fig:fptype_lat_aq_glfff} and 
Figure~\ref{fig:fptype_lat_aq_lirmhd}, respectively.
In each figure,
different field-line type is represented by bars of different colors,
as indicated.
The heights of the bars correspond to the ratios relative to the total number 
of the lines in each panel.
For the bars too short to be visible, the ratios are printed above the
respective bars.
The panels from top down show
the ratio distributions 
of All time, Before 2016 (Active Time), and After 2016 (Quiet Time).
In each panel, 
the distributions in low- ($0^\circ-30^\circ$), mid- ($30^\circ-60^\circ$)
and high- ($60^\circ-90^\circ$) latitudes are placed from left to right.
For CL10
(the field lines extending to a non-CH region),
the latitude is determined by the foot point in the coronal hole.
For CL11a and CL11n
(the field lines connecting different CHs and confined in a same CH),
the latitude is determined by the foot point at
the higher latitude.

One significant difference between the MHD and PFSS results
is that the most numerous field line type 
is OP1 in CH$_{\rm MHD}$ but is 
 usually 
CL11n in CH$_{\rm 193}$.
Despite the differences,
CL10 and CL11a, the two field line types that cross the boundaries of coronal
holes,
can be found in both PFSS and MHD magnetic fields,
with higher occurrence during active time (panel b, before 2016).
%especially in the lower latitude regions.
CL10 in the PFSS magnetic fields
is the dominant type in the low-latitude ($0^\circ-30^\circ$) range, 
and is the second most numerous in the same latitude range in the MHD results.
CL11a is more likely to be found in the low-latitude range in the MHD results
and in the mid-latitude ($30^\circ-60^\circ$) range in the PFSS results.
In both PFSS and MHD magnetic field, CL11a is the rarest type of field lines
in the coronal holes.
%and is least likely to be found in the high latitudes  during active time,
%with zero case in the PFSS and only four cases in the MHD magnetic fields.
In short, 
Figure~\ref{fig:fptype_lat_aq_glfff} and Figure~\ref{fig:fptype_lat_aq_lirmhd} 
indicate that boundary-crossing field lines, CL10 and CL11a,
are relatively more common when the Sun is more active.

\section{Discussion}
\label{sec:discuss}
\begin{figure}
  \centering
  \includegraphics[width=.8\textwidth]{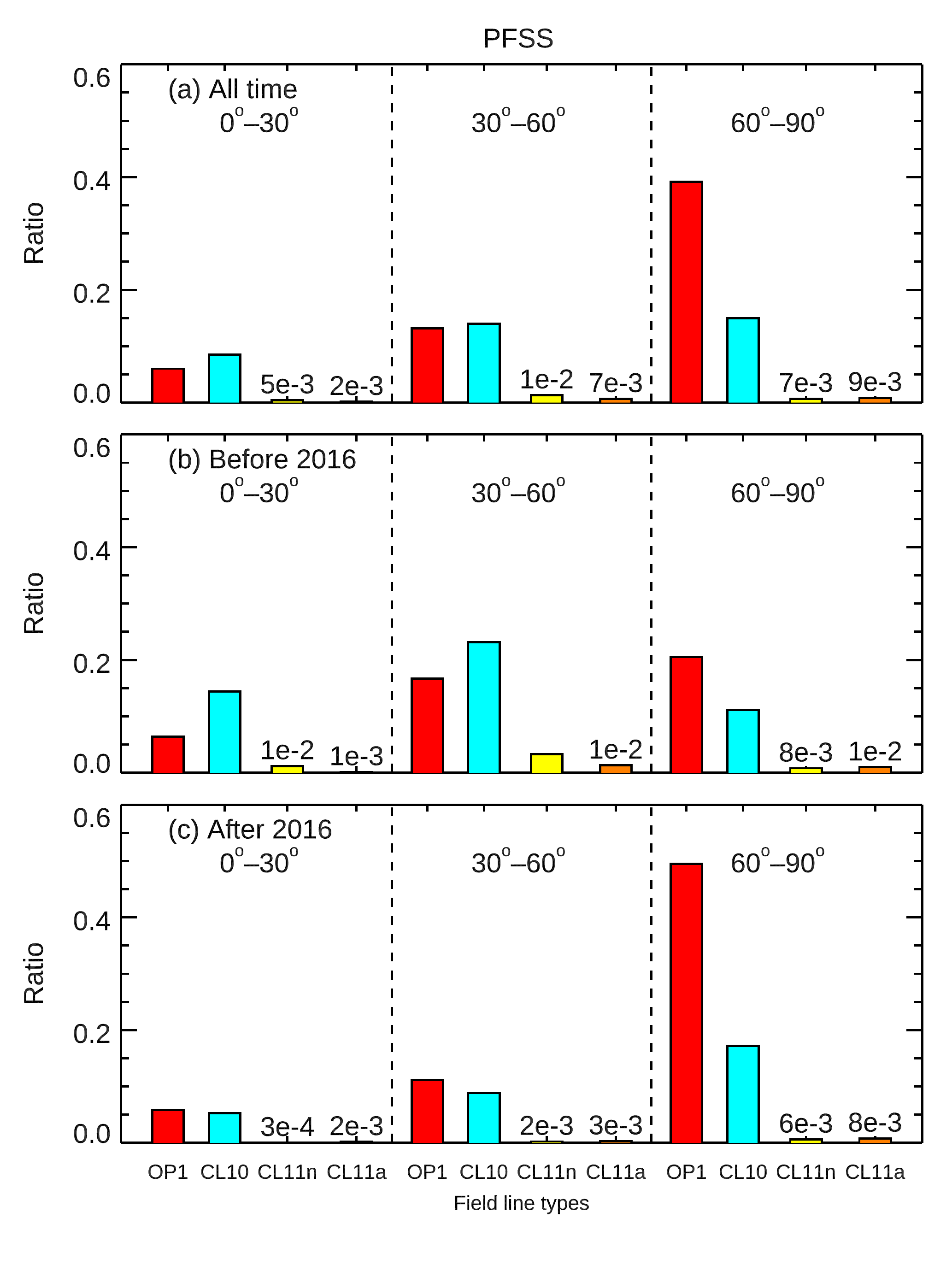}
  \caption{The ratio distribution of OP1 (red bars), CL10 (cyan bars),
           CL11n (yellow bars), and CL11a (orange bars) in different
           latitudes and during active (before 2016) and quiet (after 2016) times.
           in CH$_{193}$ with 10 degree pixel size.
           The format is same as that in Fig.~\ref{fig:fptype_lat_aq_glfff}.
          }
  \label{fig:fptype_lat_aq_10deg}
\end{figure}

The lower ratio of CL11n in the MHD coronal holes can partly be due to 
 the larger pixel size in the MHD data relative to that 
in the PFSS magnetic fields.
%Lower spatial resolution 
 Larger pixel size  can significantly reduce the number of closed field
lines inside the CHs because the bi-polar structures smaller than 
 a pixel would not be resolvable.
To test this proposition, we down-sampled the HMI synoptic maps to 
a spatial resolution of approximately 10$^\circ$, 
and re-calculated the ratio  distributions of different
field-line types. The result is plotted in Fig.~\ref{fig:fptype_lat_aq_10deg}.
The format is same as that of Fig.~\ref{fig:fptype_lat_aq_glfff}.
A comparison between Fig.~\ref{fig:fptype_lat_aq_10deg} and
Fig.~\ref{fig:fptype_lat_aq_glfff}
shows that the ratio of CL11n is significantly reduced after the down sampling.
With the reduction of CL11n, OP1 becomes the most numerous field-line type
in the high latitudes. 
In the mid- and low-latitudes, however, OP1 is still not as dominant over
other field line types as seen in CH$_{\rm MHD}$ 
(Fig.~\ref{fig:fptype_lat_aq_lirmhd}).

Another possible source for the dominance of OP1 in CH$_{\rm MHD}$ is the
the distance after which the field lines are considered open.
In this study, 
this distance is chosen to be 2.5$R_\odot$ from the Sun center.
Although the thermal pressure is mostly higher than 
the tangential magnetic pressure in the MHD model at this distance, 
it is not impossible that some closed field lines can extend higher than
this height.
In the PFSS model, 
since the source surface is the upper boundary,
changing its location can change the computed magnetic field,
thereby changing the field lines.
To check whether the relatively higher percentages of the boundary-crossing 
field lines and CHs without OP1 in the PFSS results
may be reduced by changing the source surface distance, $R_{\rm ss}$,
we adjusted $R_{\rm ss}$ from 3.0 to 1.5$R_\odot$ for
six selected Carrington Rotations, three in the active times
(CR2105, 2124, and 2136)
and three in the quiet times (CR2174, 2179, and 2183).
The results indicate that
decreasing $R_{\rm ss}$ monotonically increases
the ratio of OP1 and decreases that of the closed field lines,
especially the boundary-crossing ones,
thereby decreasing the percentage of CH$_{193}$ without OP1.
Nevertheless, 
there are still more than a quarter of CH$_{193}$ without OP1 in these six
selected Carrington Rotations
even when $R_{\rm ss}$ is set at 1.5$R_\odot$.

The differences in the approximations and assumptions
implemented in the PFSS and MHD models inevitably also 
contribute to the difference in the percentage of OP1 
between CH$_{\rm MHD}$ and CH$_{\rm 193}$, especially
in the mid- and low-latitude regions.
For instance, under the current-free restriction in the PFSS model,
neighboring points with opposite polarities would be more likely connected 
by closed field lines, thereby lowering the number of OP1.
The heating mechanisms implemented in the MHD model may
lead to brighter foot points of closed field lines and darker OP1 foot points.
As a result, the dark regions would consist of mostly open magnetic
field lines.

\begin{figure}
  \centering
  \includegraphics[width=.9\textwidth,clip]{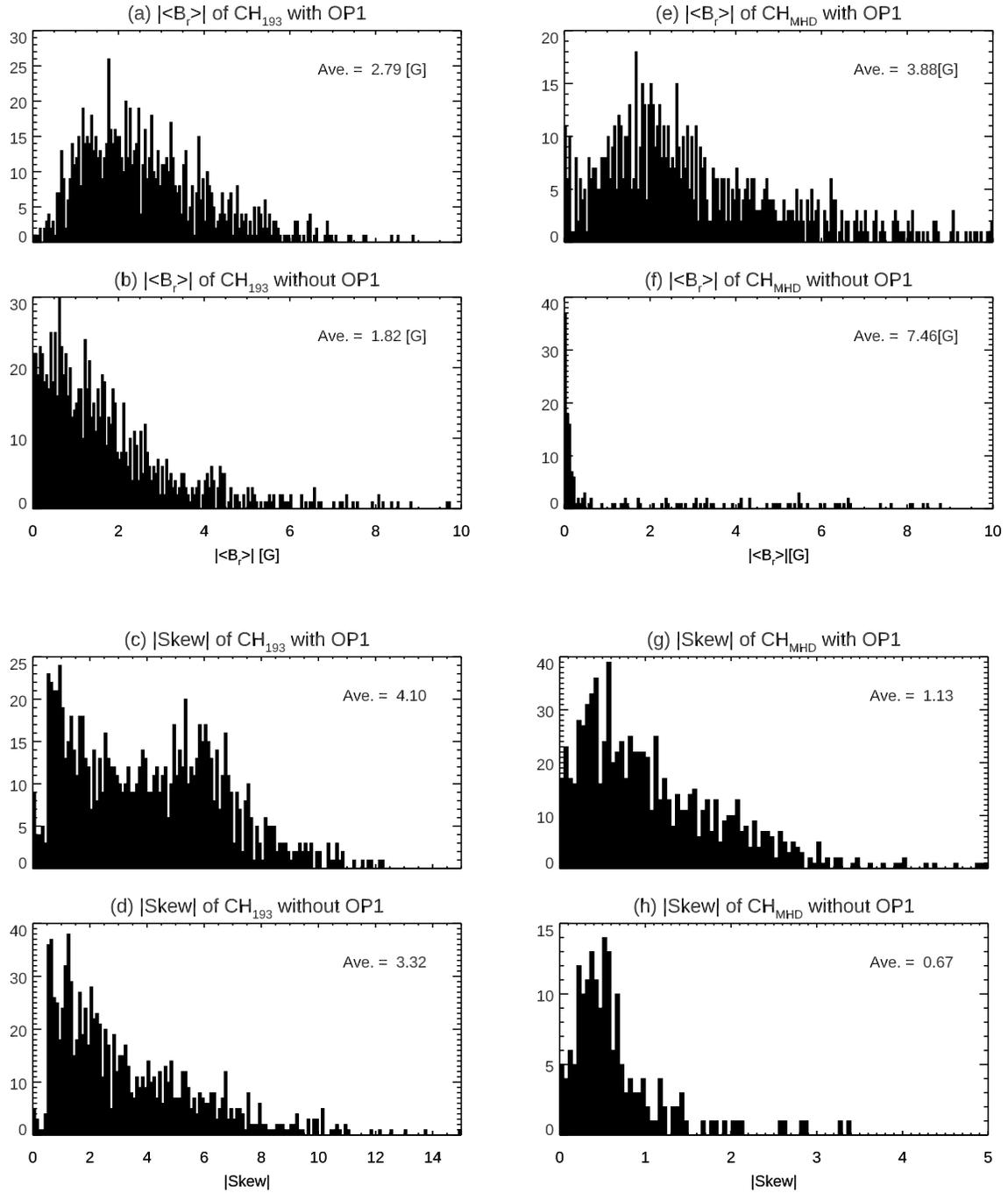}
  \caption{
            The histograms of $|\langle B_r \rangle|$ (a, b, e and f) 
            and $|{\rm Skew}|$ (c, d, g and h)
            for the coronal holes with and without open field lines (OP1),
            as indicated in the panel titles.
            The y-axis is the number of coronal holes.
            The average of each distribution is printed in the corresponding
            panel.
            The results for CH$_{193}$  and CH$_{\rm MHD}$ are placed
            in the left and right panels, respectively.
          }
  \label{fig:op1Br}
\end{figure}

\begin{figure}
  \centering
  \includegraphics[width=0.9\textwidth,clip]{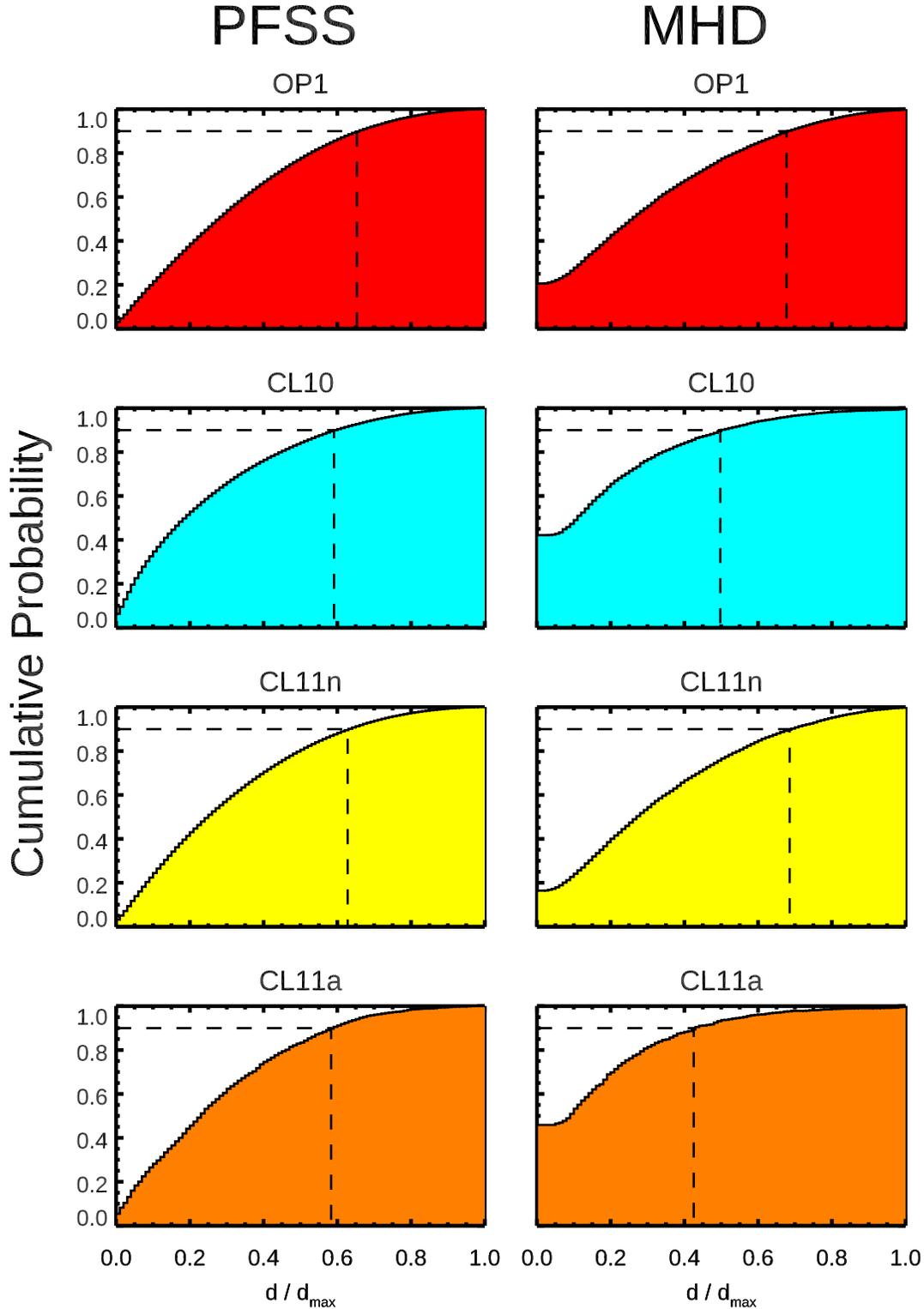}
  \caption{
           The cumulative probability of the distance of the field-line foot 
           point to the nearest boundary for different types of field lines, 
           as indicated in the titles of the corresponding panels.
           The results of CH$_{193}$  and CH$_{\rm MHD}$ are placed
           in the left and right columns.
           The distances, $d$, in a coronal hole are normalized by 
           the maximum distance, $d_{\rm max}$, in the same
           coronal hole. The x-axis is the normalized distance 
           ($d/d_{\rm max}$),
           and y-axis is the cumulative probability.
           The dashed lines mark the normalized
           distance at which the cumulative probability reaches 90\%.
          }
  \label{fig:db_sph}
\end{figure}

In addition to the percentages of CL11n and OP1, the histogram of 
$|\langle B_r \rangle|$ in the coronal holes without open magnetic fields
(Fig.~\ref{fig:op1Br}) and the cumulative probability of the distance
to the boundary (Fig.~\ref{fig:db_sph}) also reveal quantitative differences
between the magnetic fields constructed in CH$_{\rm MHD}$ and CH$_{193}$.

In Fig.~\ref{fig:op1Br}, panel f shows that 
$|\langle B_r \rangle|$
%the average magnetic field 
of most CH$_{\rm MHD}$ without OP1 is nearly zero. In other words,
their $|\langle B_r \rangle|$ is lower than
the $|\langle B_r \rangle|$ threshold of 2~G. Therefore, these
regions can only be qualified as coronal holes due to skewed distribution of
the magnetic fields.
This indicates that while the percentage of the boundary-crossing field lines
in these CH$_{\rm MHD}$
is sufficiently high to skew the distribution of the signed magnetic field,
their magnetic fields are not strong enough to shift the average $B_r$.

In Fig.~\ref{fig:db_sph}, all right panels (CH$_{\rm MHD}$) show 
a large cumulative probability 
at the zero normalized distance while the probability at the zeroth bin in
the left panels (CH$_{193}$) is very low.
This is because the pixel size in the MHD maps and magnetic fields is much
larger than that in the AIA-193 maps for the same CH area. 
Therefore, the chance of a pixel to be at the boundary is much higher
in a CH$_{\rm MHD}$ than in a CH$_{193}$.
Consequently, the probability for a field line
to be in a boundary pixel is also higher in a CH$_{\rm MHD}$ than
CH$_{193}$.

\section{Summary}
\label{sec:summary}

The coronal holes, being predominantly unipolar, are often considered to be
the regions with open magnetic field lines and, therefore, 
source regions for high-speed solar wind.
\citet{HLL2019ApJ}, however, showed that some 
coronal holes do not intersect with open magnetic field regions,
indicating that they do not contain open magnetic field lines.
The aim of our study is to explain such inconsistency
by examining the magnetic field structures of the coronal holes.
We construct the magnetic fields in the coronal holes using
PFSS model,
and four types of magnetic structures are identified:
open magnetic field line (OP1), closed magnetic field line confined
in a same coronal hole (CL11n), closed field line with the other end
extending to a bright region (CL10), and 
closed field line connecting two CHs (CL11a).
As a comparison, we use the MHD model distributed by Predictive Science
\citep{predsci}
to compute synthetic coronal holes and their magnetic structures.
All four types of magnetic structures are also found in the synthetic CHs.
Despite some quantitative differences between the magnetic fields
constructed by PFSS and MHD,
the results from both models reveal that
a significant number of coronal holes do not contain
any open magnetic field lines and that their unipolarity is due to
the closed field lines extending beyond the boundary of the coronal hole
to a bright region (CL10) or another coronal hole (CL11a),
with some CL10 connecting to an active region or ending near another
coronal hole.
Our results indicate that
these CHs without OP1 are on average smaller and less unipolar
than the CHs with open magnetic field lines.
Such coronal holes are unlikely to be the source regions of high-speed
solar winds,
and can be one reason for the low consistency level 
between CHs and high-speed solar wind source regions reported by
\citet{HLL2017NatSR}.

The statistical analysis of the boundary-crossing field lines 
indicate that a non-negligible percentage of
these field lines are not rooted in the boundary regions,
and that they are more likely to occur during the active times.
The temperature ratio of the field-line foot points
indicates higher temperature variations 
along the boundary-crossing closed field lines than along the 
non-boundary-crossing ones.

These closed field lines connecting different coronal holes
and between coronal holes and active regions may be formed by
some prior dynamics and interactions between these regions,
and may indicate some underlying large-scale magnetic structures connecting
these regions, as suggested by \citet{MOS1997SoPh, IMO1998ASPC}.
Investigation of these suppositions would require application of 
a time-dependent model to single magnetograms to
better understand the formation mechanism and the physical properties
of these field lines.

\acknowledgments

This work is supported by Ministry of Science and Technology of Taiwan under 
the grant number MOST 109-2111-M-008-002 and 
grand number MOST 109-2116-M-002-031.
We thank \href{http://www.predsci.com/}{Predictive Science} team for making their simulation data available to
public.

\appendix
\label{sec:app}

The first step of the down-sampling procedure is to apply a spherical harmonic
transformation on the HMI synoptic map:
\begin{equation}
  \hat{B}_{lm} = \int_{0}^{2\pi}\int_1^{-1}{B_r(\theta, \phi)}
  Y_{lm}^*(\theta, \phi)
  \mathrm{d}\cos\theta\mathrm{d}\phi
\end{equation}
using the routine of \texttt{SSWIDL}, where $B_r(\theta, \phi)$ is the
radial component of magnetic field at colatitude $\theta$ and longitude $\phi$,
and $Y_{lm}^*(\theta, \phi)$ is the complex conjugate of spherical harmonic
function of degree $l$ and order $m$, and $\hat{B}_{lm}$ is the harmonic
coefficients of $B_r(\theta, \phi)$.

Next, 
a Gaussian function which peaks at $l = 0, m = 0$ and reduces to half maximum 
at $\sqrt{l^2+m^2}=180$
is applied to
$\hat{B}_{lm}$:
\begin{equation}
  \hat{B}_{lm}' = \hat{B}_{lm}\exp{\left(-\frac{l^2+m^2}{2\sigma^2}\right)}
\end{equation}
where $\sigma$ is the standard deviation of the Gaussian function.

Lastly, the low-pass filtered harmonic coefficients are transformed back to
a coarser $(\theta', \phi')$ grid:
\begin{equation}
  B_r'(\theta', \phi') = \sum_{l=0}^{1440}\sum_{m=0}^{l}
  \hat{B}'_{lm}Y_{lm}(\theta', \phi')
\end{equation}
which contains 360 longitudes and 180 latitudes.
Therefore, the smoothed HMI synoptic map has a size of $360\times180$.
The inversion also uses the routine of \texttt{SSWIDL}.

\bibliography{ref}{}
\bibliographystyle{aasjournal}

\end{document}